\documentclass[fleqn,usenatbib]{mnras}

\usepackage{newtxtext,newtxmath}
\usepackage[T1]{fontenc}
\DeclareRobustCommand{\VAN}[3]{#2}
\let\VANthebibliography\thebibliography
\def\thebibliography{\DeclareRobustCommand{\VAN}[3]{##3}\VANthebibliography}

\usepackage{graphicx}	% Including figure files
\usepackage{amsmath}	% Advanced maths commands

\newcommand\bb[1]{\mbox{\boldmath{$#1$}}}

\newcommand{\vthe}{v_{\mathrm{th}e}}

\newcommand{\ompe}{\omega_{\mathrm{p}e}}
\newcommand{\vecE}{\bb{E}}
\newcommand{\vecv}{\bb{v}}
\newcommand{\vecB}{\bb{B}}
\newcommand{\hatvece}{\hat{\bb{e}}}
\newcommand{\vb}{v_\mathrm{b}}
\newcommand{\nb}{n_\mathrm{b}}
\newcommand{\rmL}{\mathrm{L}}
\newcommand{\rmT}{\mathrm{T}}
\newcommand{\rmS}{\mathrm{S}}

\title[Plasma Emission in Fully Kinetic Simulations]{Fundamental, Harmonic, and Third-harmonic Plasma Emission from Beam-plasma Instabilities: A First-principles Precursor for Astrophysical Radio Bursts}
\author[F.\ Bacchini \& A.~A.\ Philippov]{
Fabio Bacchini$^{1,2}$\thanks{E-mail: fabio.bacchini@kuleuven.be} and
Alexander A.\ Philippov$^{3}$
\\
$^{1}$Centre for mathematical Plasma Astrophysics, Department of Mathematics, KU Leuven, Celestijnenlaan 200B, B-3001 Leuven, Belgium\\
$^{2}$Royal Belgian Institute for Space Aeronomy, Solar-Terrestrial Centre of Excellence, Ringlaan 3, 1180 Uccle, Belgium\\
$^{3}$Department of Physics, University of Maryland, College Park, MD 20742, USA
}

\begin{document}
\label{firstpage}
\pagerange{\pageref{firstpage}--\pageref{lastpage}}
\maketitle

\begin{abstract}
Electromagnetic fundamental and harmonic emission is ubiquitously observed throughout the heliosphere, and in particular it is commonly associated with the occurrence of Type II and III solar radio bursts. Classical analytic calculations for the plasma-emission process, though useful, are limited to idealized situations; a conclusive numerical verification of this theory is still lacking, with earlier studies often providing contradicting results on e.g.\ the precise parameter space in which fundamental and harmonic emission can be produced. To accurately capture the chain of mechanisms underlying plasma emission --- from precursor plasma processes to the generation of electromagnetic waves over long times --- we perform large-scale, first-principles simulations of beam-plasma instabilities. By employing a very large number of computational particles we achieve very low numerical noise, and explore (with an array of simulations) a wide parameter space determined by the beam-plasma density ratio and the ion-to-electron temperature ratio. In particular, we observe direct evidence of both fundamental and harmonic plasma emission when the beam-to-background density ratio $\le0.005$ (with beam-to-background energy ratio $\sim 0.5$), tightly constraining this threshold. We observe that, asymptotically, in this regime $\sim0.1\%$ of the initial beam energy is converted into harmonic emission, and $\sim0.001\%$ into fundamental emission. In contrast with previous studies, we also find that this emission is independent of the ion-to-electron temperature ratio. In addition, we report the direct detection of third-harmonic emission in all of our simulations, at power levels compatible with observations. Our findings have important consequences for understanding the viable conditions leading to plasma emission in space systems, and for the interpretation of observed electromagnetic signals throughout the heliosphere.
\end{abstract}


%%%%%%%%%%%%%%%%%%%%%%%%%%%%%%%%%%%%%%%%%%%%%%%%%%

%%%%%%%%%%%%%%%%% BODY OF PAPER %%%%%%%%%%%%%%%%%%

\section{Introduction}
\label{sec:intro}

The production of electromagnetic waves is frequently observed during energetic outbursts from the solar surface, e.g.\ as Type II and III radio bursts (e.g.\ \citealt{dulk1985,stanislavsky2022}). Such observations allow us to probe the plasma conditions of solar eruptive events (e.g.\  \citealt{wildsmerd1972,reidratcliffe2014,ndacyayisenga2023} and references therein), and commonly report the presence of waves at a frequency equal to and/or double the local electron plasma frequency $\ompe=\sqrt{4\pi |q_e|^2 n_e/m_e}$ (where $q_e$ and $m_e$ are the electron charge and mass, and $n_e$ is the local electron density). Despite decades of research, however, it is still rather unclear how exactly such ``fundamental'' (at $\omega\sim\ompe$) and ``harmonic'' (at $\omega=2\ompe$) emission originates from collective plasma mechanisms.

Theoretical work on the subject spans several decades and has converged on the widely accepted three-wave-interaction model (\citealt{ginzburgzhelezniakov1958,melrose1970b,melrose1970a,zheleznyakovzaitsev1970,melrose2017}). The latter considers a two-stage process initiated by the excitation of forward-propagating Langmuir ($\rmL$) waves, e.g.\ via the nonlinear evolution of a precursor electrostatic instability. Fundamental emission can then directly originate from the decay of $\rmL$ waves (found along the Langmuir dispersion curve, $\omega^2=\ompe^2+3k^2\vthe^2$, with $\vthe$ the electron thermal speed) into ion acoustic waves (IAWs) {(labeled $\rmS_1$)} and electromagnetic waves ($\rmT_1$, found along the plasma dispersion curve $\omega^2=\ompe^2+k^2c^2$, where $c$ is the speed of light) at approximately the local $\ompe$. Forward-propagating electrostatic waves can also {coalesce with, or decay into, IAWs (labeled $\rmS_2$, which are distinct from $\rmS_1$)} to produce backward-propagating ($\rmL'$) modes; the subsequent $\rmL\mbox{--}\rmL'$ interaction results in harmonic electromagnetic emission ($\rmT_2$ waves). The whole process can be summarized as
\begin{equation*}
\begin{aligned}
    \rmL\to\rmT_1+\rmS_1 & \quad \mbox{(Fundamental emission)} \\
    \left\{\begin{matrix}
    \rmL\pm\rmS_2\to\rmL' \\
    \rmL+\rmL'\to\rmT_2
    \end{matrix}\right.
    & \quad \mbox{(Harmonic emission)}
\end{aligned}
\end{equation*}
and this three-wave interaction satisfies conservation of momentum and energy in the weak-turbulence limit (e.g.\ \citealt{tsytovich1972}). Hence, each step of the process directly translates into requency and wavenumber sums.

Despite general agreement on the mechanism originating plasma emission, no direct numerical simulation of the whole process (starting from the initial instability to the nonlinear wave-wave interaction) has provided conclusive results supporting analytic calculations. This is mainly due to the tremendous computational resources involved: fully kinetic simulations are necessary, with large (ion-scale) system sizes evolved for long times (much longer than the instability relaxation time) while fully resolving the electron physics. Additionally, the modes involved are expected to grow and saturate only up to a small fraction of the input kinetic energy, implying that numerical noise must be kept very low. Previous works attempting such simulations have focused on fully kinetic Particle-in-Cell (PIC) methods applied to the interaction of dilute electron beams with a cold plasma background, which ideally initiates the plasma-emission process (e.g.\ \citealt{cairnsrobinson1998}). These simulations are very demanding, and involve specific choices of the ion-to-electron temperature ratio and very low beam-to-background density ratios. The latter results in very long quasilinear relaxation times and large computational costs. A low density ratio is however absolutely necessary: the Langmuir dispersion relation can be severely modified by the beam when the density ratio is $\gtrsim10^{-4}$ (see \citealt{cairns1989}), producing unstable modes at frequencies $\omega_\rmL$ {substantially different from} the electron plasma frequency $\ompe$. This is detrimental for plasma emission, since these modified $\rmL$ waves cannot participate in the $\rmL\pm\rmS_2\to\rmL'$ process discussed above due to a violation of \emph{matching conditions}: $\rmS_2$ waves cannot exist at frequencies {much larger or smaller} than the corresponding IAW frequency $\omega_{\rmS_2}$, and therefore they cannot compensate for the {smaller or larger} $\omega_\rmL$ obtained in this case. Low density ratios are therefore essential to maintain the coupling between beam and Langmuir unstable modes at (approximately) $\ompe$, resulting in electrostatic $\rmL$ fluctuations that can initiate the three-wave interaction.

In simulations, harmonic emission has been reportedly observed in one- and two-dimensional setups with a single electron beam, multiple counterstreaming beams, and with and without background magnetic fields (e.g.\ \citealt{kasaba2001,sakai2005,umeda2010,tsiklauri2011,thurgoodtsiklauri2015,thurgoodtsiklauri2016,henri2019,lee2019,krafftsavoini2022a,lee2022,lazar2023}).
\cite{thurgoodtsiklauri2015} provided a first decisive demonstration of plasma emission in numerical calculations. That work showed that the possibility of plasma emission is contingent upon the frequency of the initial electrostatic waves generated by the beam-plasma instability, and that these waves may be prohibited from participating in the necessary three-wave interactions due to frequency conservation requirements. However, strong evidence for clearly distinguishable fundamental emission (expected to arise with much smaller power than the harmonic signal) was not ubiquitously detected. In addition, the precise plasma conditions (particularly the threshold beam-to-background density ratio and ion-to-electron temperature ratio) under which harmonic emission occurs have yet to be conclusively identified in simulations.
In essence, a thorough quantification of the saturated energy level of different modes and of the parameter space in which the three-wave interaction occurs is still missing. This information is however not only fundamental to understand the origin of commonly observed radio signals in space, but also for experiments attempting to reproduce space-plasma conditions (e.g.\ \citealt{marques2020} and references therein). We also note that, during the development of this work, a first conclusive detection of fundamental modes in PIC simulations of beam-plasma systems was reported by \cite{zhang2022}, providing numerical evidence to support the three-wave-interaction model of plasma emission.

In this Letter we build on past groundwork by performing an array of PIC simulations of beam-plasma interaction, with the aim to observe and quantify the subsequent plasma emission and the saturated energy levels of each mode involved in the process. {Our multidimensional simulations are of very large size\footnote{Only surpassed in domain size by the slightly larger run presented in \cite{krafftsavoini2022a}.} and unprecedented duration}, to accurately capture the nonlinear evolution of the system over long time scales and to comfortably fit all the wave modes involved in the mechanisms of interest; to obtain converged and reliable results, in our runs we achieve very low levels of numerical noise by employing large numbers of particles per cell. With this approach, we explore the dependence of plasma emission on the initial plasma conditions by varying the beam-to-background density ratio and the ion-to-electron temperature ratio. This parameter scan allows us to converge on the quantification of plasma-emission processes directly applicable to specific astrophysical situations, particularly for unmagnetized systems with freely streaming electron beams.

\section{Numerical model and parameters}
\label{sec:params}

We perform two-dimensional, high-resolution PIC simulations with \textsc{TRISTAN-MP} (\citealt{buneman1993,spitkovsky2005,hakobyan2023zeltron}). Our numerical setup consists of a square periodic box of size $L\times L$, where we initialize a single electron beam with bulk velocity $\vecv_\mathrm{b}=\vb\hatvece_x$, a background thermal electron population with near-zero (see below) mean velocity, and a thermal ion population with mass ratio $m_i/m_e=1836$. The beam and background electrons are assigned different numerical weights to achieve a specific beam-to-background density ratio $\alpha=\nb/n_0$. To impose charge neutrality at $t=0$, the ion density is set to $n_i=\nb+n_0$. Initial current neutrality is obtained by adding a small bulk velocity $\vecv_0=v_0\hatvece_x$ to the background electron population, such that $\nb \vb + n_0 v_0=0$. Finally, each electron species is initialized with a specific thermal speed $\vthe$ to achieve a chosen ratio $\vb/\vthe$; ions are assigned a different temperature based on a specific choice of the ion-to-electron temperature ratio $T_i/T_e$. The free parameters in the simulation are thus $\alpha=\nb/n_0$, $\vb/\vthe$, and $T_i/T_e$, as well as the system size $L$ and final simulation time $t_\mathrm{f}$.

\begin{table*}
\centering
\begin{tabular}{cccccccccc} % four columns, alignment for each
\hline
Run & $\alpha$ & $T_i/T_e$ & $L/L_{\rmS_2}$ & $L/L_{\rmS_1}$ & $L/L_\mathrm{k}$ & $L/L_{\rmT_2}$ & $L/L_{\rmT_1}$ & $\Gamma_\mathrm{k}$ & $\Gamma_\mathrm{QL}$ \\
\hline
Reference & 0.005 & 0.1 & 53.34 & 26.67 & 26.67 & 23.10 & 2.31 & 0.85 & 0.25 \\
\hline
DR1 & 0.1 & 0.1 & 50.99 & 25.49 & 25.49 & 22.08 & 2.21 & 0.31 & 5 \\
\hline
DR2 & 0.025 & 0.1 & 52.82 & 26.41 & 26.41 & 22.87 & 2.29 & 0.5 & 1.25 \\
\hline
DR3 & 0.001 & 0.1 & 53.45 & 26.72 & 26.72 & 23.14 & 2.31 & 1.45 & 0.05 \\
\hline
TR1 & 0.005 & 1 & 53.34 & 26.67 & 26.67 & 23.10 & 2.31 & 0.85 & 0.25 \\
\hline
TR2 & 0.005 & 0.66 & 53.34 & 26.67 & 26.67 & 23.10 & 2.31 & 0.85 & 0.25 \\
\hline
TR3 & 0.005 & 0.33 & 53.34 & 26.67 & 26.67 & 23.10 & 2.31 & 0.85 & 0.25 \\
\hline
\end{tabular}
\caption{In all cases, $kT_e/(m_ec)=0.0025$, $\vb/\vthe=10$, and $m_i/m_e=1836$; $L=84 c/\ompe$ and the grid spacing $\Delta x = (1/32) c/\ompe$. We always employ 2560 particles per cell for each species (beam electrons, background electrons, background ions).}
\label{tab:param}
\end{table*}

To choose the free parameters in our simulations, we consider the following requirements:
\begin{itemize}
    \item Transition of the initial beam-plasma instability to the \textit{kinetic regime}\footnote{Note that our Eq.~\eqref{eq:kinregime} expresses exactly the same scaling shown in Eq.~(7) of \cite{cairns1989}.} (e.g.\ \citealt{oneilmalmberg1968,cairns1989}): we impose that
    \begin{equation}
    k_{\|}\vthe\gtrsim\gamma_\mathrm{bp}=\ompe\frac{\sqrt{3}}{2^{4/3}}\alpha^{1/3},
    \label{eq:kinregime}
    \end{equation}
    where $\gamma_\mathrm{bp}$ is the maximum growth rate of the beam-plasma instability in the cold (i.e.\ fluid) limit. We are therefore demanding that the transit time $1/(k_\|\vthe)$ of the beam particles over one wavelength (where the most-unstable wavenumber is $k_{\|}=\ompe/\vb$) be shorter than the typical fluid-instability growth time $1/\gamma_\mathrm{bp}$. In this regime, a significant fraction of the beam particles cannot interact with the waves excited in the cold beam regime. In terms of our simulation parameters, we thus search for the condition
    \begin{equation}
        \Gamma_\mathrm{k} = \frac{\vthe/\vb}{\sqrt{3}\alpha^{1/3}/2^{4/3}} \gtrsim 1.
    \end{equation}
    To fit the fastest-growing instability wavelength into the simulation box, we also require $L>L_\mathrm{k}=2\pi/(\ompe/\vb)$.

    \item {Existence of $\rmT_1$ waves: the wavelength of fundamental emission can be found by using conservation of momentum and energy for the $\rmL\to\rmT_1+\rmS_1$ decay (e.g.\ \citealt{cairns1989}), i.e.\ $\omega_\rmL=\omega_{\rmS_1}+\omega_{\rmT_1}$ and $\bb{k}_\rmL=\bb{k}_{\rmS_1}+\bb{k}_{\rmT_1}$. From the latter, since we know that $k_{\rmT_1}\ll k_{\rmS_1}$, we obtain $k_{\rmS_1}\simeq k_\rmL\simeq k_\|=\ompe/v_b$. Furthermore, from a Taylor expansion of the dispersion curves, we have the $\rmL$ and $\rmT_1$ frequencies $\omega_\rmL\simeq\ompe[1+3k_\rmL^2\vthe^2/(2\ompe^2)]$ and $\omega_{\rmT_1}\simeq\ompe[1+k_{\rmT_1}^2c^2/(2\ompe^2)]$. The IAW frequency is given by the approximate IAW dispersion $\omega\simeq k\vthe\sqrt{m_e/m_i}\sqrt{1+3T_i/T_e}$; knowing $k_{\rmS_1}$, we can find $\omega_{\rmS_1}$ and therefore $\omega_{\rmT_1}$. Finally, in the limit $m_e/m_i\ll1$, we obtain the $\rmT_1$ wavenumber $k_{\rmT_1}\simeq\sqrt{3}\ompe\vthe/(\vb c)$. Therefore, we search for $L>L_{\rmT_1}=2\pi/(\sqrt{3}\ompe\vthe/(\vb c))$, such that $\rmT_1$ waves fit into the simulation box.}

    \item {Existence of $\rmT_2$ waves: knowing the $\rmT_2$ frequency $\omega_{\rmT_2}=2\ompe$, from the plasma dispersion curve $\omega_{\rmT_2}^2=\ompe^2+k_{\rmT_2}^2c^2$ we find the corresponding wavenumber, $k_{\rmT_2}=\sqrt{3}\ompe/c$. We thus require the simulation box to fit the wavelength of harmonic emission, i.e.\ we search for $L>L_{\rmT_2}=2\pi/(\sqrt{3}\ompe/c)$. Note that since $\rmL'$ is a background mode with frequency $\omega_{\rmL'}\simeq\ompe$, conservation of energy also gives the $\rmS_2$ frequency $\omega_{\rmS_2}\simeq|\omega_\rmL-\ompe|$.}

    \item {Existence of $\rmS$ waves: the simulation box must fit the wavelength of $\rmS_1$ and $\rmS_2$ IAWs. The former have wavenumber $k_{\rmS_1}\simeq\ompe/\vb$ (see above). The $\rmS_2$ wavenumber can be found by applying conservation of momentum to the $\rmL\to\rmL'+\rmS_2$ mechanism (i.e.\ $\bb{k}_\rmL=\bb{k}_{\rmL'}+\bb{k}_{\rmS_2}$), which gives $k_{\rmS_2}\simeq2\ompe/\vb$. To fit both types of IAWs in the simulation box, we search for $L>L_{\rmS_1}=2\pi/(\ompe/\vb)$ and $L>L_{\rmS_2}=2\pi/(2\ompe/\vb)$ (hence, $\rmS_1$ waves guide the choice of $L$). In addition, IAWs can only develop if $T_i/T_e$ is sufficiently low due to otherwise strong Landau damping and subsequent vanishing of the IAW growth rate (e.g.\ \citealt{friedgould1961}). This imposes an upper limit on the admissible ion temperature (see discussion below and in Section~\ref{sec:results}).}

    \item Regime of \textit{weak turbulence}: for the validity of quasilinear relaxation, we require that the beam energy be much smaller than the background thermal energy, i.e.
    \begin{equation}
        \alpha\ll2\left(\frac{\vthe}{\vb}\right)^2,
    \end{equation}
    or equivalently $\Gamma_{\mathrm{QL}} = \alpha/(2\vthe^2/\vb^2)\ll 1$.
    Taking into account high-order corrections in the Langmuir dispersion gives even stringent conditions, which however do not appear to affect our results\footnote{Demanding that nonlinear corrections to the Langmuir frequency be much smaller than thermal corrections gives $\alpha \ll3(\vthe/\vb)^4$, i.e.\ a much more stringent condition on the simulation parameters. Although this requirement is not respected in our runs, we do not find evidence that our results are significantly affected by this violation.}.
\end{itemize}
These considerations thus identify the following requirements: (i) beam-to-background density and velocity ratios such that $\Gamma_\mathrm{k}\sim1$ and also $\Gamma_\mathrm{QL}\ll1$; (ii) a simulation box that is larger than the largest among the wavelengths of interest, i.e.
\begin{equation}
    L_{\rmS_2}<L_{\rmS_1}\sim L_\mathrm{k}<L_{\rmT_2}<L_{\rmT_1}<L,
\end{equation}
whose ordering is valid when $\vb/\vthe\gg1$ and $\alpha\ll1$; and (iii) in principle, an ion-to-electron temperature ratio that is small enough to allow for the production of IAWs.

It is not straightforward to choose simulation parameters that respect all constraints. From the length-scale ordering above, it is clear that we require $L>L_\mathrm{\rmT_1}$, since we want to capture fundamental emission\footnote{We note that the runs presented in \cite{thurgoodtsiklauri2015} do not employ a system size that can fit $\rmT_1$ modes, which likely explains their observation of a lack of fundamental emission. In addition, their run \#2 does not respect $\Gamma_\mathrm{QL}\ll1$.}. However, a compromise must be made when setting $\vthe/\vb$ and $\alpha$: increasing or decreasing one of the two correspondingly increases or decreases $\Gamma_\mathrm{k}$ and $\Gamma_\mathrm{QL}$, potentially violating one of our constraints. Moreover, $\alpha$ cannot be decreased arbitrarily, mainly because to observe plasma emission we need to evolve the system over time scales much larger than the instability quasilinear time $t_\mathrm{QL}=\ompe^{-1}(\vb/\vthe)^2/\alpha$ (where we can assume $\vthe\sim\vb$ after the transition to kinetic regime). With smaller $\alpha$, simulations thus become increasingly more expensive; in these long simulations, PIC codes (in particular those employing explicit schemes) will also accumulate larger numerical errors in the energy, progressively invalidating the results. A compromise must then be found between affordable computational costs, numerical accuracy, and appropriate values of $\alpha$ and $\vthe/\vb$ producing physically interesting results. A final point of interest stems from the need, in principle, to ensure the development of IAWs to achieve harmonic emission via three-wave interaction: the growth rate of the IAW mode indeed decreases with the increase of the temperature ratio $T_i/T_e$ due to electron Landau damping (e.g.\ \citealt{friedgould1961}).

Bearing in mind all these constraints, we set up a reference simulation with the following parameters: $\vb/\vthe=10$, $\alpha=\alpha_\mathrm{ref}=0.005$, $L=84c/\ompe$, $T_i/T_e=0.1$ ({with the normalized electron temperature $kT_e/(m_e/c^2)/c=0.0025$ and $\vthe/c=\sqrt{kT_e/m_e}/c=0.05$}, see Table~\ref{tab:param}). With this choice, we marginally respect all guidelines detailed above.
We are subsequently interested in studying the effect of varying $\alpha$ and $T_i/T_e$ on the plasma-emission process, to quantify the parameter thresholds at which the three-wave interaction is impeded. This parameter-space exploration is motivated by the fact that, at fixed $\vb/\vthe$ and $L>L_{\rmT_1}$ (to ensure that all wavelengths of interest fit into the box), the beam-to-background density ratio $\alpha$ is the determining factor to control how well most other constraints are respected. In particular, it can be observed that smaller $\alpha$ values translate into better achieving the kinetic and weak-turbulence regimes. In addition, exploring a range of $T_i/T_e$ allows us to verify the expectation that a reduced temperature ratio is a necessary condition to obtain harmonic emission (\citealt{thurgoodtsiklauri2015,zhang2022}). We will in fact show that substantial $\rmT_2$ emission can be measured even when $T_i/T_e$ is large.

In the next Sections we first present and analyze in detail the reference run; we then discuss two series of simulations, where we explore the effect of varying $\alpha$ and $T_i/T_e$. Our runs are summarized in Table~\ref{tab:param}. With this parameter scan, we assess how the process of plasma emission depends on the underlying plasma conditions, and we determine which simulations can be reliably employed to draw conclusions on this mechanism.

\begin{figure*}
\centering
\includegraphics[width=1\textwidth, trim={0mm 77.5mm 15mm 0mm}, clip]{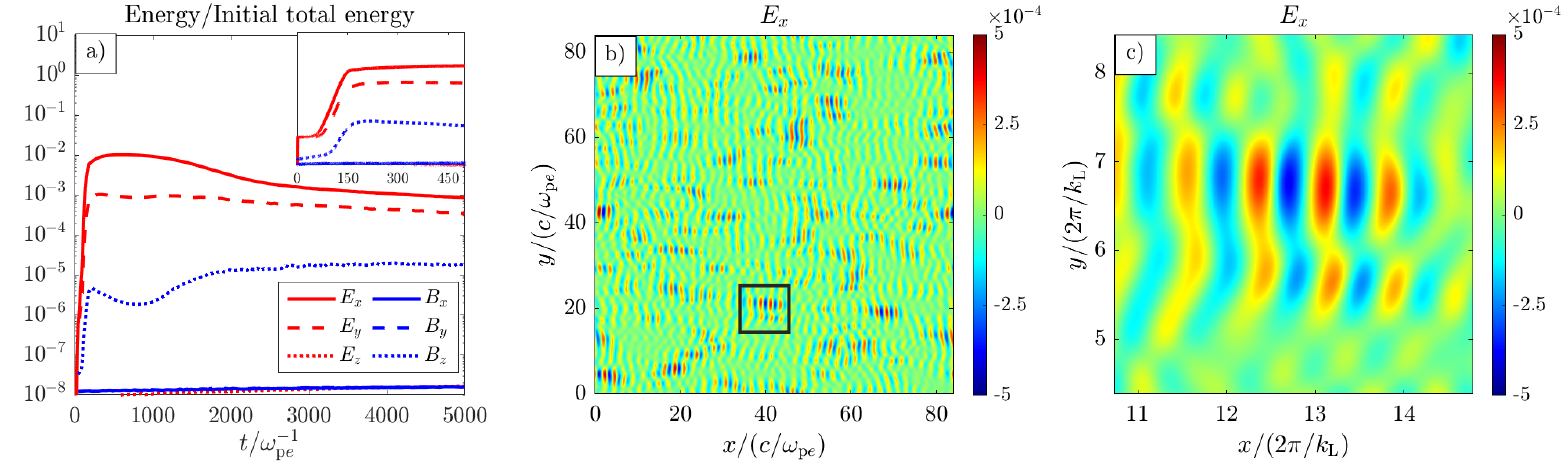}
\caption{Diagnostics for the reference simulation in Table~\ref{tab:param}. Panel \textit{a}: evolution in time of the electromagnetic energy in each electric- and magnetic-field component. The inset in the top-right corner is a zoomed-in view of $t\in[0,500]\ompe^{-1}$, showing the development of the initial beam-plasma instability. Panel \textit{b}: spatial distribution of the electrostatic field $E_x$ at $t=500\ompe^{-1}$. Panel \textit{c}: zoomed-in view onto the region inside the black rectangle of panel \textit{b}. The wave structures in this region can be identified as $\rmL$ modes.}
\label{fig:energy_ex}
\end{figure*}

\begin{figure*}
\centering
\includegraphics[width=1\textwidth, trim={0mm 0mm 15mm 0mm}, clip]{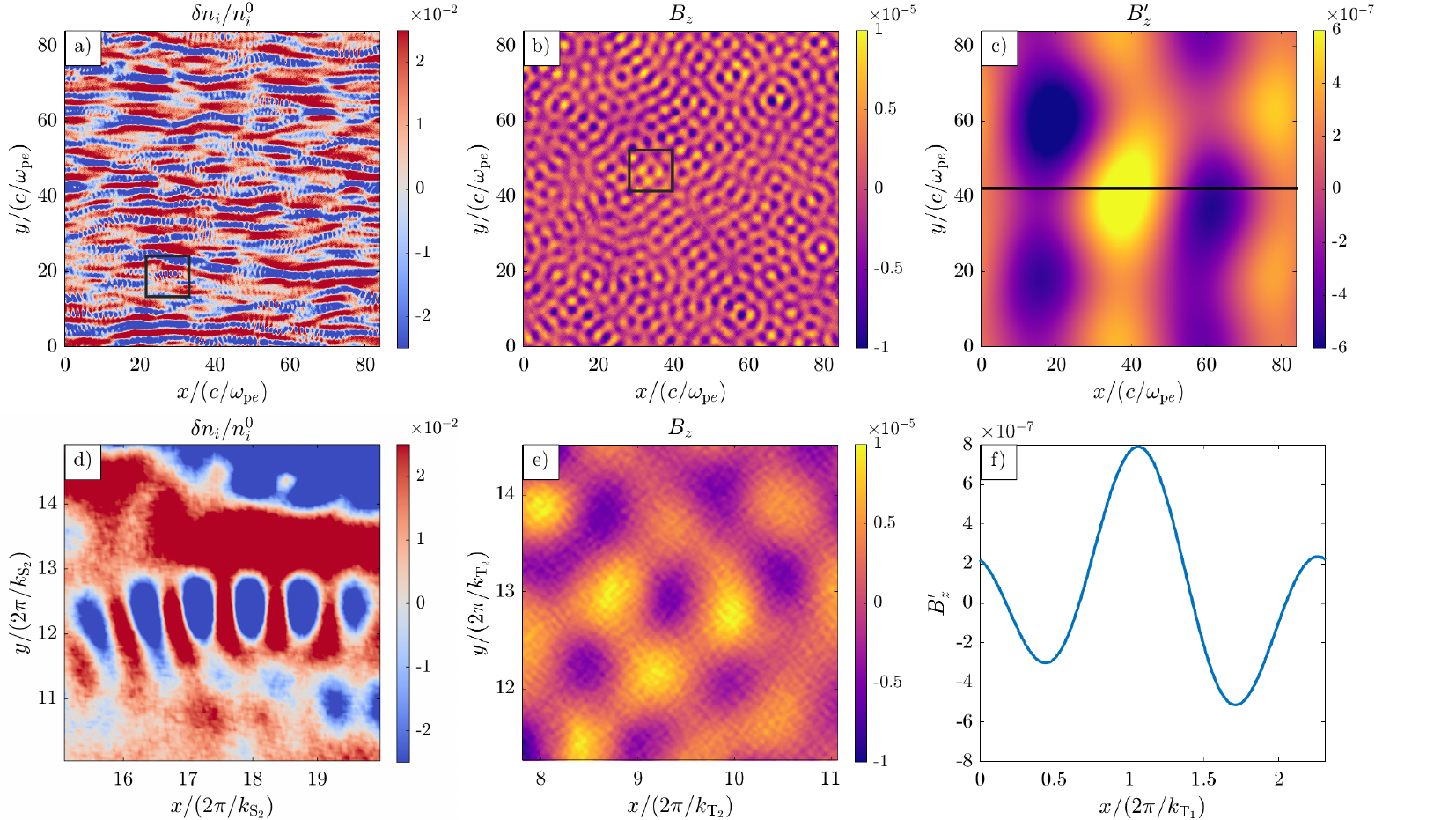}
\caption{Diagnostics for the reference simulation in Table~\ref{tab:param}. Panels \textit{a} and \textit{d}: spatial distribution of ion-density fluctuations at $t=1500\ompe^{-1}$ and zoomed-in view onto the region marked by the black rectangle, showing a wave structure identified as an $\rmS_2$ mode. Similarly for panels \textit{b} and \textit{e}: distribution of the electromagnetic field $B_z$ and zoomed-in view onto the wave structure of a $\rmT_2$ mode. Panel \textit{c}: distribution of the filtered magnetic field $B'_z$ obtained by removing spectral modes with $k>k_{\rmT_1}$ from $B_z$. Large-scale wave structures can be observed in the filtered field. Panel \textit{f}: a cut along $y=L/2$ (black line in panel \textit{c}) shows that the visible waves in $B'_z$ can be identified as $\rmT_1$ modes.}
\label{fig:dni_bz}
\end{figure*}

\section{Reference run and dependence of plasma emission on density ratio}
\label{sec:results}

In this Section, we analyze the plasma-emission mechanism by considering our reference run (see Table~\ref{tab:param}) as well as a varying density ratio $\alpha$. Our reference simulation employs $\vb/\vthe=10$, $\alpha=\alpha_\mathrm{ref}=0.005$, $T_i/T_e=0.1$, and $L=84 c/\ompe$. Our numerical grid has resolution $\Delta x=(1/32)c/\ompe$ (i.e.\ $2688^2$ cells), with 2560 particles per cell for each particle species (beam electrons, background electrons, background ions)\footnote{We find that such a large number number of particles per cell is absolutely necessary to combat numerical noise and reach convergence: we verified that our results are quantitatively unchanged when initializing $\ge1280$ particles per cell. Alternative approaches, such as the Delta-f method (e.g.\ \citealt{sydora2003}) could be considered to combat numerical noise without the need for a large number of computational particles}.

\subsection{Reference Run: Development of Electrostatic and Electromagnetic Modes}

In Fig.~\ref{fig:energy_ex} (panel \textit{a}) we show the time history of the volume-averaged energy in each component of the electric and magnetic field ($\vecE$ and $\vecB$, respectively). The system's evolution is divided in two distinct phases: during $t\in[0,500]\ompe^{-1}$ (blow-up inset in panel \textit{a}), we observe the development of the initial electrostatic beam-plasma instability, which feeds off the input beam energy reservoir. This phase corresponds to the excitation of Langmuir waves\footnote{{Note that during this initial phase, $B_z$ also grows in energy simply as a result of the beam-plasma instability without yet developing coherent electromagnetic waves at multiples of $\ompe$.}} appearing in the spatial distribution of the electrostatic field $E_x$, shown at $t=500\ompe^{-1}$ in panel \textit{b} of the same Figure. Spatial fluctuations in $E_x$ can be observed throughout the domain; their wavelength is compatible with $2\pi/k_\rmL$ (where $k_\rmL=\ompe/\vb$), which identifies them as $\rmL$ modes.

Mode conversion of forward-propagating $\rmL$ waves occurs between $t=1000\ompe^{-1}$ and $t=2000\ompe^{-1}$: in Fig.~\ref{fig:dni_bz} (panel \textit{a} with a zoomed-in view in panel \textit{d}), we observe the presence, at $t=2000\ompe^{-1}$, of spatial fluctuations in the ion density corresponding to the $\rmS_2$ IAW wavelength $2\pi/k_{\rmS_2}$ (where $k_{\rmS_2}\simeq2k_\rmL$); simultaneously, the electromagnetic (out-of-plane) field $B_z$ develops coherent fluctuations that can be observed (panel \textit{b} with a zoomed-in view in panel \textit{e}) to be compatible with the wavelength of harmonic emission $2\pi/k_{\rmT_2}$ (where $k_{\rmT_2}=\sqrt{3}\ompe/c$). Moreover, filtering out high-frequency waves reveals underlying wave structures: by removing spectral modes with $k>k_{\rmT_1}=\sqrt{3}(\ompe/c)(\vthe/\vb)$, the filtered field $B'_z$ (panel \textit{c} and cut-in view in panel \textit{f}) shows large-scale fluctuations that can be identified as $\rmT_1$ modes.

Our results provide one among the few clear, direct identifications of waves produced via fundamental and harmonic emission in a fully kinetic simulation of beam-plasma interaction. The detection of these modes confirms that the criteria outlined in Section~\ref{sec:params} produce the expected three-wave mechanism efficiently, accurately capturing the dynamics of the corresponding mode conversion. During the development of this work, \cite{zhang2022} also showed a similar result, and the analysis presented here broadly agrees with theirs. In the following Sections, we significantly expand this investigation by analyzing more quantitatively the development of all waves involved in the plasma-emission process and the dependence on the physical parameters employed.

\subsection{Reference Run: Spectral Power in the Three-wave Modes}
\begin{figure*}
\centering
\includegraphics[width=1\textwidth, trim={0mm 0mm 15mm 0mm}, clip]{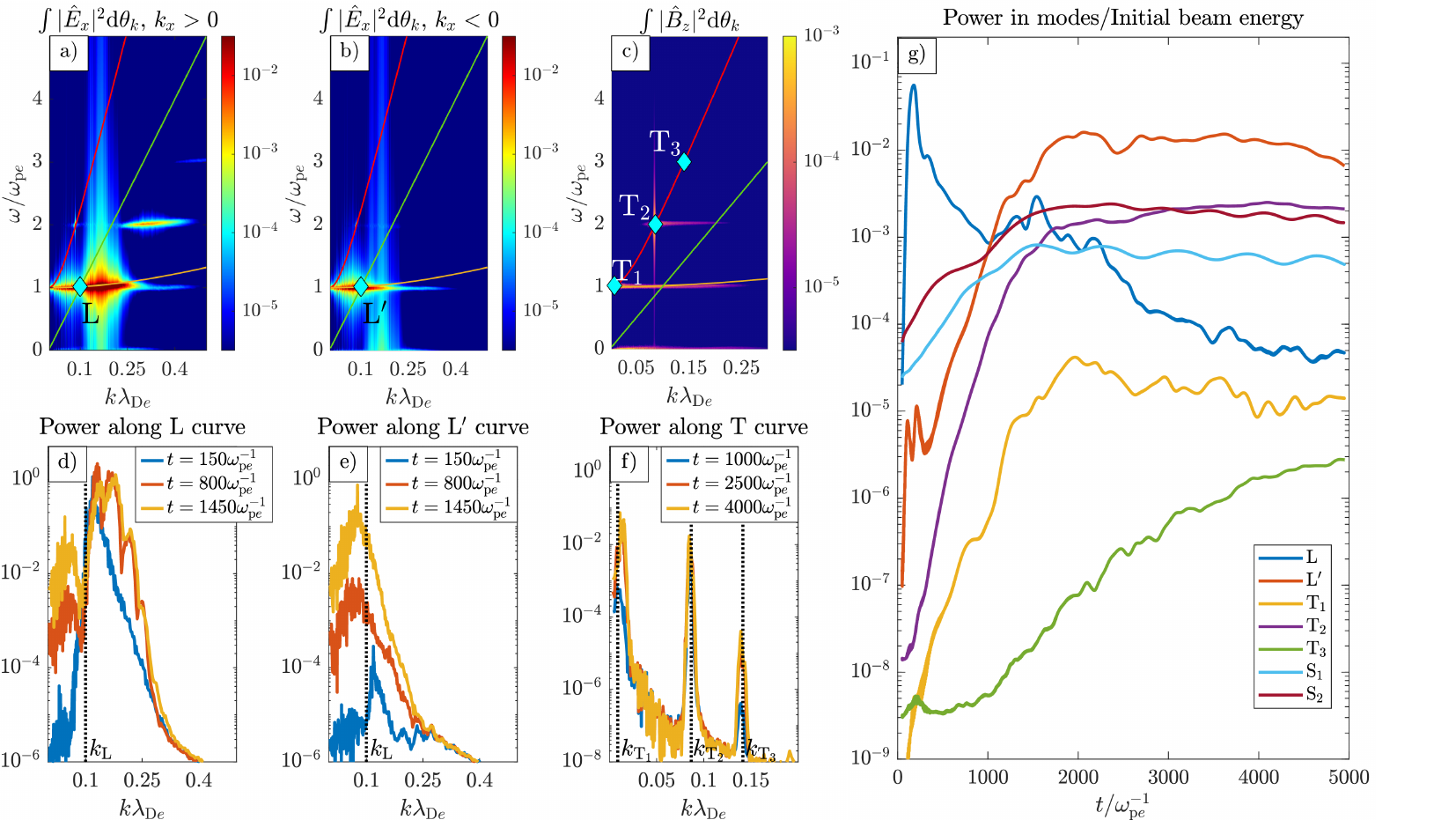}
\caption{FFT diagnostics for the reference simulation in Table \ref{tab:param}. Panels \textit{a}, \textit{b}, and \textit{c}: space-time FFTs of $E_x$ (with time window $t\in[750,1250]\ompe^{-1}$) and $B_z$ (with time window $t\in[500,1500]\ompe^{-1}$). The FFTs are cylindrically integrated in the $k_xk_y$-plane; for $E_x$, we perform the integration separately for the half-planes $k_x>0$ and $k_x<0$. The solid lines indicate the dispersion curves for the Langmuir mode (orange), the beam mode (green), and the plasma oscillation mode (red). Cyan diamonds indicate the $(k,\omega)$-locations of the electrostatic and electromagnetic modes involved in the three-wave interaction. Panels \textit{d}, \textit{e}, and \textit{f}: spectral power in $E_x$ and $B_z$ measured at subsequent times, respectively along the Langmuir dispersion curve (separately for $k_x>0$ and $k_x<0$) and the plasma dispersion curve. Dashed vertical lines indicate the wavenumber of $\rmL$ and $\rmT$ modes. Panel \textit{g}: evolution in time of the spectral power (normalized to the initial beam energy) in $E_x$, $B_z$, and ion-density fluctuations measured at the $(k,\omega)$-locations of all modes of interest.}
\label{fig:fft}
\end{figure*}

In Fig.~\ref{fig:fft} we show the spectral analysis of the modes of interest for the reference run. Panels \textit{a} and \textit{b} show the isotropic (i.e.\ cylindrically integrated in the $k_xk_y$-plane, via $\theta_k=\tan^{-1}(k_y/k_x)$) space-time FFT of the electrostatic $E_x$ at $t\in[500,1500]\ompe^{-1}$. The integration is performed separately for forward- ($k_x>0$) and backward-propagating ($k_x<0$) fluctuations, in order to diagnose counterpropagating $\rmL$ and $\rmL'$ modes respectively. The cyan diamond in these plots indicates the phase-space location $(k_\rmL,\omega_\rmL)$ of the $\rmL$ mode at the intersection of the Langmuir dispersion curve $\omega=\sqrt{\ompe^2+3k^2 \vthe^2}$ (orange line) with the beam dispersion curve $\omega=k \vb$ (green line).
% We observe in particular (as expected of nonlinear Landau damping of electrostatic fluctuations) that both $\rmL$ and $\rmL'$ modes tend to broaden and drift toward larger wavenumbers over time.
Panel \textit{c} similarly shows the isotropic space-time FFT of the magnetic field $B_z$ for $t\in[500,1500]\ompe^{-1}$, with the location of $\rmT$ modes indicated by cyan diamonds along the plasma dispersion curve $\omega=\sqrt{\ompe^2+k^2c^2}$ (red line). The spectral power in electrostatic and electromagnetic modes along the corresponding dispersion curves is shown at subsequent times in panels \textit{d}, \textit{e}, and \textit{f}; in all cases, the spectra are clearly peaked around the wavenumbers of the modes involved in the three-wave interaction (dashed vertical lines), i.e.\ $\rmL$, $\rmT_1$, and $\rmT_2$, with the power in each mode increasing over time before stabilizing around a saturated value.

Interestingly, we note that an additional peak of spectral power arises at $(k_{\rmT_3},\omega_{\rmT_3})=(\sqrt{8}\ompe/c,3\ompe)$ along the plasma dispersion curve. This ``third-harmonic'' emission ($\rmT_3$) is weaker than both harmonic and fundamental signals, but it is clearly detectable as a byproduct of the three-wave interaction process, as we will demonstrate later. Higher-harmonic emission has been previously studied in fully kinetic simulations with a single parameter set (e.g.\ \citealt{rhee2009,krafftsavoini2022b}). We discuss this third-harmonic emission process more extensively in the next Sections.

The evolution in time of the peak power in each mode of interest (measured at the corresponding $(k,\omega)$), including $\rmS$ modes from ion-density fluctuations, is plotted in panel \textit{g} of Fig.~\ref{fig:fft}. We can observe that the evolution of the peak power in the forward-propagating $\rmL$ modes is clearly correlated with the energetics of the primary beam-plasma instability that jumpstarts the plasma-emission process: power in the $\rmL$ waves sharply rises within the first $100\ompe^{-1}$, reaching a maximum immediately followed by a slower decrease. During this time, $\rmS$ modes also gain power as ion-density fluctuations develop. The subsequent dynamics builds on the nonlinear stage of the primary beam-plasma mode: power in $\rmL$ modes decreases and backward-propagating $\rmL'$ modes develop together with $\rmT_2$, $\rmS_1$, and $\rmS_2$ modes, reaching saturated values at $t\simeq1500\ompe^{-1}$. This is a clear indication that, at the very least, the $\rmL+\rmL'\to\rmT_2$ process is actively pumping energy from the electrostatic modes into electromagnetic radiation. The resulting harmonic emission saturates at $\sim0.1\%$ of the initial beam energy. In terms of fundamental emission, we observe that $\rmT_1$ modes also steadily gain power and saturate over the same time scales of harmonic emission, but reach considerably lower saturation levels. This suggests that the conversion of electrostatic fluctuations into large-scale $\rmT_1$ waves is less efficient, with the latter reaching saturation at around $0.001\%$ of the initial beam energy.

Finally, measuring the power deposited into third-harmonic electromagnetic waves reveals that $\rmT_3$ modes also gain power and saturate, although over longer time scales and at much smaller energies than all other modes. Third-harmonic modes indeed receive only a small ($\sim10^{-6}$) fraction of the initial beam energy, which is however sufficient to clearly distinguish $\rmT_3$ emission from the background noise. In the next Sections, we will show that in all our experiments third-harmonic waves invariably arise.

\subsection{Effect of Density Ratio}

\begin{figure*}
\centering
\includegraphics[width=1\textwidth, trim={0mm 10mm 20mm 0mm}, clip]{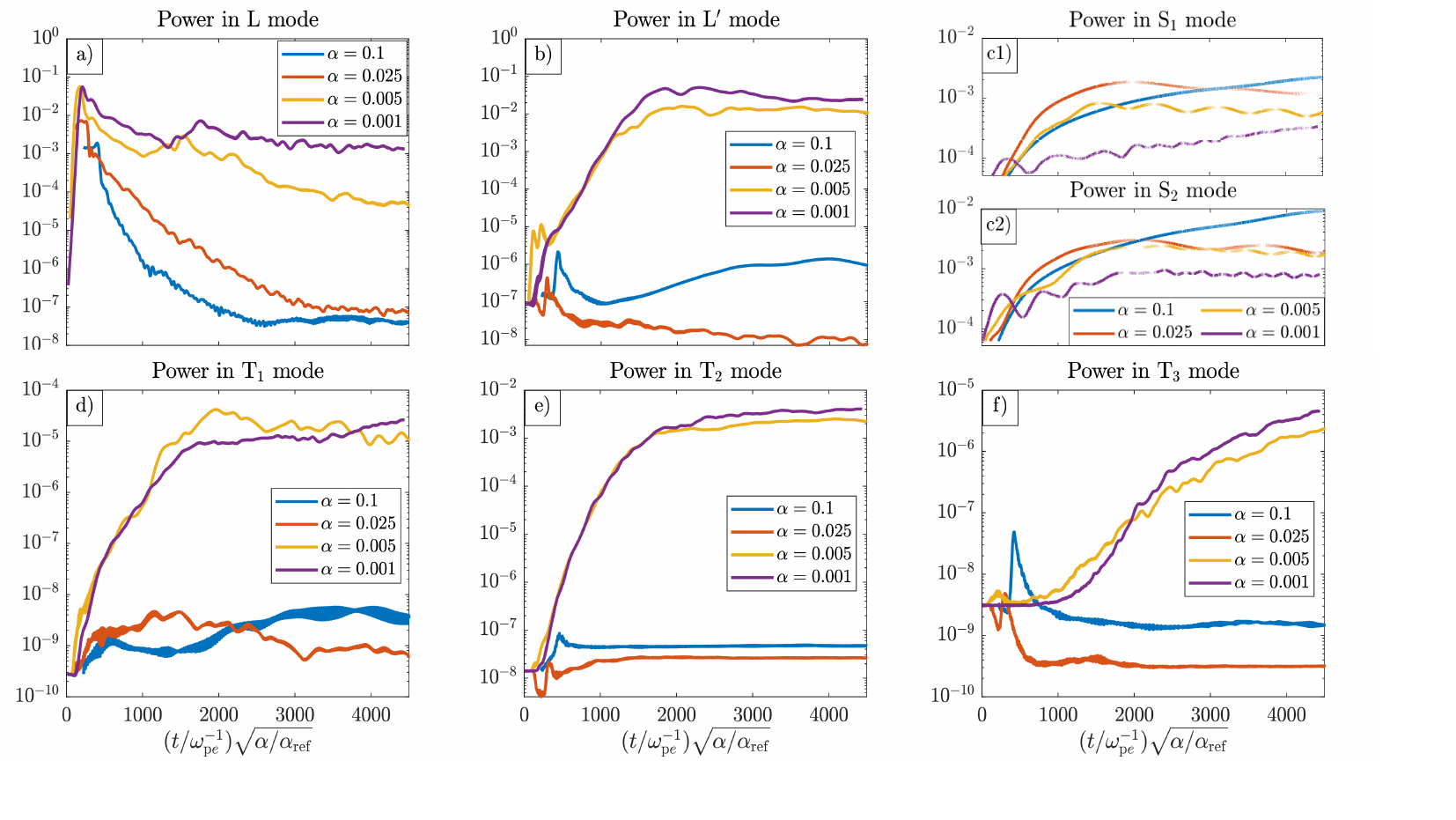}
\caption{Evolution in time of the spectral power (normalized to the initial beam energy) measured at the $(k,\omega)$-location of the modes involved in the three-wave interaction process, for the reference run in Table~\ref{tab:param} ($\alpha_\mathrm{ref}=0.005$) compared with runs DR1, DR2, DR3. Time is scaled by a factor $\sqrt{\alpha/\alpha_\mathrm{ref}}$ to account for the different instability time scales imposed by the different density ratio $\alpha$.}
\label{fig:density_ratio}
\end{figure*}

Fig.~\ref{fig:density_ratio} shows the comparison of spectral power in the different modes (including $\rmT_3$) over time, for variable initial beam-to-background density ratios $\alpha=0.1,0.025,0.005,0.001$ (with $\alpha_\mathrm{ref}=0.005$ for the reference simulation; all other parameters are kept fixed across runs, see Table~\ref{tab:param}). For each run, time is scaled by a factor $\sqrt{\alpha/\alpha_\mathrm{ref}}$ to account for the different time scales of the beam-plasma instability that initiates the system's evolution. As discussed in the previous sections, a large density ratio is expected not to result in the production of a strong $\rmT_2$ signal due to the violation of matching conditions. Here, we explicitly observe the suppression of electromagnetic-wave production as $\alpha$ decreases.

While the production of $\rmL$ waves (panel \textit{a}) by the initial beam-plasma instability is detected in all cases (albeit with different dynamics for large $\alpha$), we observe that for $\alpha>0.005$ the subsequent three-wave interaction is not triggered. Indeed, we measure significant power in backward-propagating $\rmL'$ waves (see panel \textit{b}) only when $\alpha\le0.005$, even though IAWs are developing in all runs (panels \textit{c}1 and \textit{c}2). The presence of counterpropagating $\rmL$ waves {at wavenumbers satisfying the three-wave interaction conditions} is in principle sufficient to ensure that $\rmT_2$ emission occurs; we indeed observe growth and saturation of power in $\rmT_2$ for the $\alpha\le0.005$ cases, and no such dynamics for larger density ratios (panel \textit{e}). Similarly, a clear, saturated $\rmT_1$ signal (roughly 100 times weaker than the $\rmT_2$ signal) develops for the low-density runs (panel \textit{e}). $\rmT_1$ emission was not detected clearly in some previous works, particularly when the system size was not taken large enough to fit $\rmT_1$ wavelengths (\citealt{thurgoodtsiklauri2015}; see Section~\ref{sec:params}).

\begin{figure*}
\centering
\includegraphics[width=1\textwidth, trim={0mm 10mm 20mm 0mm}, clip]{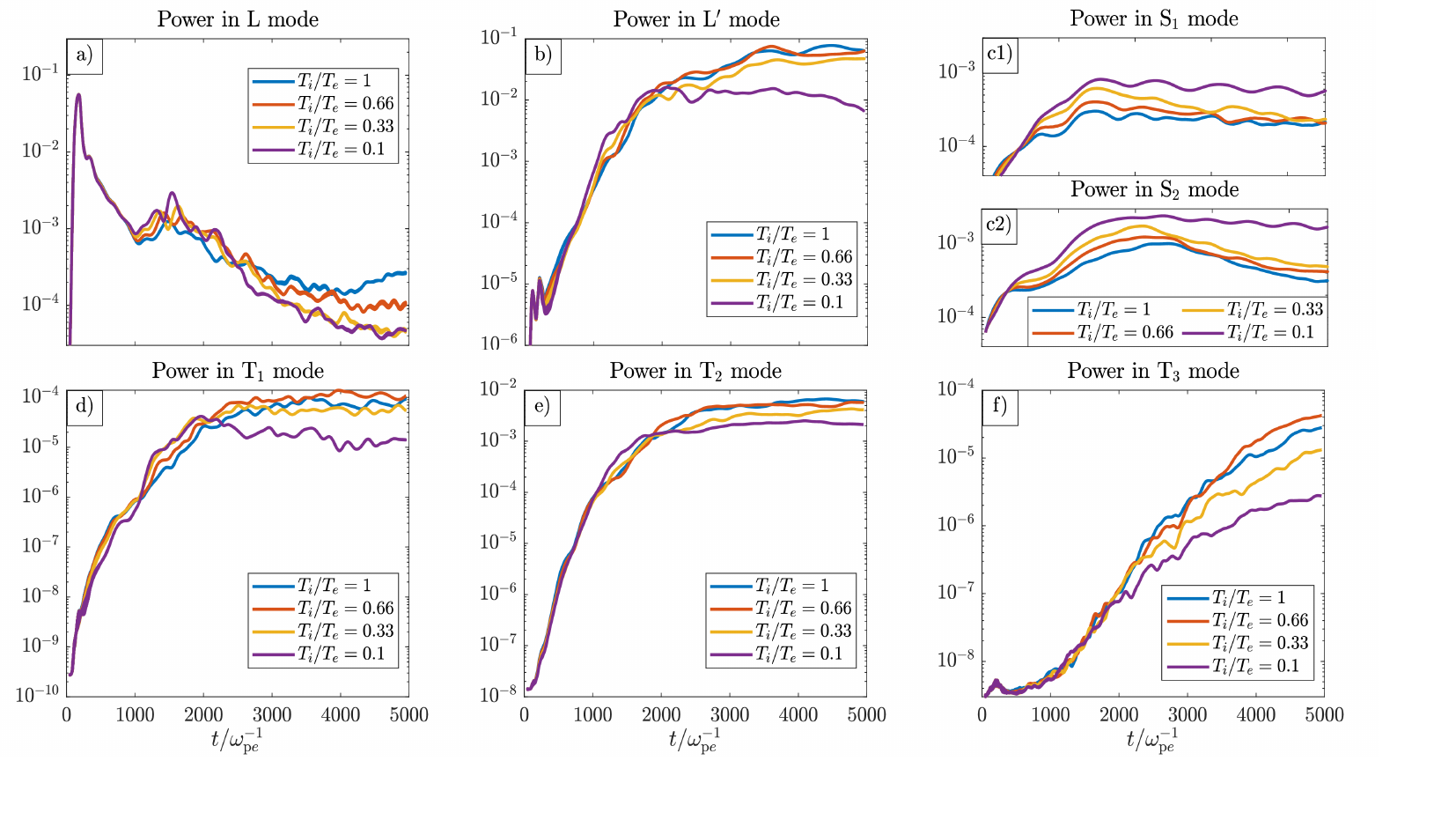}
\caption{Evolution in time of the spectral power (normalized to the initial beam energy) measured at the $(k,\omega)$-location of the modes involved in the three-wave interaction process, for the reference run in Table~\ref{tab:param} ($T_i/T_e=0.1$) compared with runs TR1, TR2, TR3.}
\label{fig:temperature_ratio}
\end{figure*}

Our results provide a definitive confirmation that both fundamental and harmonic plasma emission can occur as a consequence of a simple beam-plasma interaction; but to observe this mechanism in simulations, it is necessary to perform large calculations of sufficient size and with sufficiently low beam-plasma density ratio. Power in the harmonic emission appears to asymptotically (for small enough $\alpha$) saturate and converge to a fraction ($\sim0.1\mbox{--}0.4\%$) of the initial beam energy; power in the fundamental emission saturates to energies roughly 100 times smaller. The fact that harmonic and fundamental emission do not occur for large $\alpha$, even though technically all basic ingredients ($\rmL$ and $\rmS$ waves) are present, is likely due to the violation in the matching conditions introduced in Section~\ref{sec:intro}. This will be further discussed in Section~\ref{sec:conclusions}.

Finally, we quantitatively confirm the presence of a third-harmonic emission type in Fig.~\ref{fig:density_ratio} (panel \textit{f}). Power in $\rmT_3$ waves clearly rises and saturates similarly to that of other electromagnetic modes, although with a visible delay and at significantly lower power levels. It is interesting to note that this third-harmonic emission only appears with a sufficiently low $\alpha$, suggesting that the mechanism for $\rmT_3$ production may require conditions similar to those allowing harmonic and fundamental emission. We will discuss such mechanisms in Section~\ref{sec:conclusions}.

\section{Dependence of plasma emission on temperature ratio}

We conclude our parameter-space exploration by comparing the reference run from Table~\ref{tab:param} ($T_i/T_e=0.1$) with a series of runs where we vary $T_i/T_e=1,0.66,0.33$. The results are shown in Fig.~\ref{fig:temperature_ratio}, again in terms of the spectral power in individual modes measured at the corresponding $(k,\omega)$ locations from the cylindrically integrated FFTs of the field quantities.

Qualitatively, we observe no drastic change in the system's evolution as the ion-to-electron temperature ratio varies between 0.1 and 1: the reference run attains saturated power levels in $\rmS$ and $\rmT$ waves moderately lower than the other runs (up to a factor $\sim10$), but overall we observe growth and saturation of power in all modes. The only notable difference arises in the evolution of IAWs (panels \textit{c}1 and \textit{c2} in Fig.~\ref{fig:temperature_ratio}): for the reference $T_i/T_e=0.1$ run, we observe the power in $\rmS_2$ modes growing and saturating, while for the other runs we only see an initial growth in spectral power followed by a steady decay. This is a signature that the production of IAWs is much less efficient for large temperature ratios, due to strong Landau damping, in accordance with linear theory (\citealt{friedgould1961}). It is then rather counterintuitive that the production of $\rmL'$ modes (panel \textit{b}), and thus of fundamental and harmonic emission (panels \textit{d} and \textit{e}), remains active even when $\rmS$ waves are scarcely excited. We also notice that the growth rate of the power in $\rmL'$ modes during $t\in[500,2000]\ompe^{-1}$ is different for runs with different $T_i/T_e$, and is larger when the temperature ratio is smaller. In runs with large $T_i/T_e$, the creation of $\rmL'$ waves can be explained by the direct backscattering of $\rmL$ waves off moving ions (e.g.\ \citealt{tsytovichkaplan1969,zlotnik1998}), which is a fundamentally different phenomenon from the collision of $\rmL$ with $\rmS$ waves that can occur when the temperature ratio is small. We discuss this aspect more in detail in Section~\ref{sec:conclusions}.

We conclude by noting that also for this parameter-space exploration we invariably measure a clear growth in the power associated with $\rmT_3$ modes. This supports the idea that third-harmonic emission occurs under the same conditions that allow for harmonic and fundamental modes, regardless of the presence (or absence) of saturated $\rmS$ modes.

\section{Discussion and conclusions}
\label{sec:conclusions}

In this Letter, we presented conclusive evidence for the efficient production of fundamental and harmonic plasma emission ($\rmT_1$ and $\rmT_2$ waves) in fully kinetic simulations of beam-plasma interactions. By performing an array of runs with different initial parameters, we observed that harmonic emission saturates asymptotically to $\sim0.1\mbox{--}0.4\%$ of the initial beam energy, while fundamental emission is roughly 100 times weaker. The emission process broadly aligns with theoretical expectations --- i.e., it is only detected when the beam-to-background density ratio $\alpha$ (as well as the energy ratio $\alpha(\vb/\vthe)^2$) is sufficiently small --- but a few of our key results are novel and require further discussion.

First, our series of simulations with variable density ratio shows that, in all cases (also for large $\alpha$), Langmuir ($\rmL$) and ion acoustic ($\rmS$) waves are produced from the precursor electrostatic instability excited by our initial conditions. According to linear theory, $\rmL$ and $\rmS$ modes are in principle the building blocks for the production of backward-propagating $\rmL'$ (at first) and $\rmT$ (as a second-order effect) waves, and thus it is not obvious why plasma emission would not occur when $\alpha>0.005$ (i.e.\ the beam energy is $\ge50\%$ of the background energy). As shown by \citealt{cairns1989}, when $\alpha\gtrsim10^{-4}$ the power in $\rmL$ modes peaks at a frequency significantly below ($\sim0.1\ompe$) that of the Langmuir dispersion curve. As a consequence, matching conditions in the frequency allowing for the interaction of $\rmL$ and $\rmS$ waves are broken ($\rmS$ modes do not exist at frequencies that can compensate the mismatch) and backward-propagating electrostatic modes are not produced, inhibiting the subsequent plasma emission. This fact was also considered by \cite{thurgoodtsiklauri2015} and \cite{zhang2022}, and we now provide solid evidence for its occurrence in a series of runs where $\alpha$ varies by two orders of magnitude. In our $\alpha<0.005$ runs (relatively close to the $\alpha\sim10^{-4}$ limit calculated by \citealt{cairns1989}), $\rmL$ modes firmly sit on the Langmuir dispersion curve and matching conditions are met, allowing for plasma emission.

A second point requiring clarification concerns our runs with variable $T_i/T_e$. In all these simulations, plasma emission is invariably detected, even though IAWs are less efficiently produced for $T_i/T_e>0.1$, preventing the {interaction of $\rmL$ and $\rmS$ waves} that results in $\rmL'$ and then $\rmT$ waves. The fact that across all runs (even for $T_i/T_e=1$) we measure comparable saturated power levels for $\rmL'$ and $\rmT$ modes can thus be explained by considering direct backscattering off moving ions. In the absence of $\rmS$ waves, a forward-propagating $\rmL$ wave can directly collide with an ion moving in the opposite direction (e.g.\ \citealt{tsytovichkaplan1969}): the electrostatic wave interacts with the surrounding cloud of electrons, which reemits the wave while transferring momentum to the ion. This causes a recoil in the ion motion and the emission of a backward-propagating $\rmL'$ wave. Although further work is required to provide more evidence linking our simulations with this mechanism, these results firmly establish that plasma emission is essentially insensitive of the ion-to-electron temperature ratio, and is instead uniquely dependent on the beam-to-background energy ratio (in our study, expressed by the density ratio). This is in contrast with previous {multidimensional} simulation studies where values $T_i/T_e\gtrsim0.1$ were assumed to produce no plasma emission. {We note that analytic works and one-dimensional simulations (e.g.\ \citealt{rha2013} and references therein) have mentioned the possibility of producing plasma emission at $T_i/T_e\sim1$ due to highly efficient ion scattering compensating for the lack of interaction with $\rmS$ waves.}

As an additional result, we emphasize that in all our runs where plasma emission occurs we also invariably detected a clear, peaked signal at $(k_{\rmT_3},\omega_{\rmT_3})=(\sqrt{8}\ompe/c,3\ompe)$, indicating the presence of third-harmonic modes along the plasma dispersion curve. Analytic calculations have suggested that interactions $\rmT_2+\rmL' \to\rmT_3$ occurring as a byproduct of harmonic emission can give rise to this third-harmonic signal (e.g.\ \citealt{zlotnik1998}); a second possibility is the direct coalescence of multiple electrostatic modes, i.e.\ $\rmL+\rmL'+\rmL\to\rmT_3$ (\citealt{kliem1992}). However, the latter is more likely for slow electron beams coupled with strong electron density gradients (\citealt{yi2007}), which does not correspond to the conditions of our simulations.
Higher-harmonic plasma emission has been observed in the past in PIC simulations {(in the case of a homogeneous-density background which we employ)} considering a single set of parameters: \cite{rhee2009} and \cite{krafftsavoini2022b} have reported the detection of third-harmonic (and higher) emission, relating it to the \cite{zlotnik1998} mechanism of $\rmT_2+\rmL$ coalescence. Our array of runs explores a broad range of simulation parameters (in terms of beam-to-background density and ion-to-electron temperature ratios) with uniform background density, where we detected third-harmonic emission in all cases where $\rmT_2$ emission occurred. Our results thus appear to support the $\rmT_2+\rmL'\to\rmT_3$ hypothesis, implying that the growth of third-harmonic modes may be feeding off the harmonic emission process.
The power we measure in third-harmonic emission is roughly 100--1000 times smaller than the power in harmonic modes, which appears to align with observations of Type II (e.g.\ \citealt{kliem1992,zlotnik1998}) and III (e.g.\ \citealt{takakurayousef1974,reinermacdowall2019}) radio bursts; still, a broader array of data (numerical and experimental) would be required to firmly establish this correspondence. {Even though more in-depth investigation is needed, our results provide the unambiguous direct detection of third-harmonic plasma emission in a large array of kinetic simulations of beam-plasma interaction which explore very different parameter regimes. This allows us to claim that third-harmonic emission could be an invariable byproduct of the three-wave-interaction process even when physical conditions vary considerably.}

Our work demonstrates efficient production of coherent emission in a relatively simple setting where a single beam interacts with a uniform background, leaving ample ground for further developments. In particular, it would be important to consider the effect of background magnetic fields and/or density inhomogeneities, as well as the presence of counterstreaming beams (\citealt{tsiklauri2011,thurgoodtsiklauri2015,thurgoodtsiklauri2016,lazar2023}). Furthermore, it is known that the emission efficiency for different modes can have strong dependence on other parameters such as heliocentric distance, background-wind speed, and ambient-density fluctuations (e.g.\ \citealt{robinsoncairns1998a,robinsoncairns1998b,robinsoncairns1998c,voshchepynets2015,krafftsavoini2021,krafftsavoini2022a}). Another possibility is the production of plasma emission without three-wave interaction when a magnetized electron beam encounters a density shear (e.g.\ \citealt{schmitztsiklauri2013}). In all these cases, a rigorous parameter-space exploration akin to the one performed here would reveal under which conditions coherent emission can be expected, relating directly to various astrophysical scenarios where electron beams are present, such as coronal loops, shock waves from coronal mass ejections, etc.

\section*{Acknowledgements}
F.B.\ would like to thank Immanuel Christopher Jebaraj and Marian Lazar for useful discussions and suggestions throughout the development of this project.
F.B.\ acknowledges support from the FED-tWIN programme (profile Prf-2020-004, project ``ENERGY'') issued by BELSPO, and from the FWO Junior Research Project G020224N granted by the Research Foundation -- Flanders (FWO).
The simulations were performed on the NSF Frontera supercomputer \citep{stanzione2020} at the Texas Advanced Computing Center (TACC, \href{http://www.tacc.utexas.edu}{www.tacc.utexas.edu}) under grant AST21006, and at the VSC (Flemish Supercomputer Center), funded by the Research Foundation Flanders (FWO) and the Flemish Government – department EWI.
This work was performed in part at Aspen Center for Physics, which is supported by National Science Foundation grant PHY-1607611.

%%%%%%%%%%%%%%%%%%%%%%%%%%%%%%%%%%%%%%%%%%%%%%%%%%
\section*{Data Availability}
The data underlying this article will be shared on reasonable request to the corresponding author.



\bibliographystyle{mnras}
% \bibliography{mylib4} % if your bibtex file is called example.bib

%\begin{thebibliography}{99}
%\end{thebibliography}

% %%%%%%%%%%%%%%%%%%%%%%%%%%%%%%%%%%%%%%%%%%%%%%%%%%

% %%%%%%%%%%%%%%%%% APPENDICES %%%%%%%%%%%%%%%%%%%%%

% \appendix

% \section{Some extra material}

% If you want to present additional material which would interrupt the flow of the main paper,
% it can be placed in an Appendix which appears after the list of references.

% %%%%%%%%%%%%%%%%%%%%%%%%%%%%%%%%%%%%%%%%%%%%%%%%%%

% Don't change these lines
\bsp	% typesetting comment
\label{lastpage}
\end{document}